\documentclass[12pt]{article}
\usepackage{amsmath,amssymb,amsbsy,amsfonts,amsthm,latexsym,amsopn,amstext,
            amsxtra,euscript,amscd}
\title{{\large\bf Algorithm for factoring some RSA and Rabin moduli}}
\author{Omar Khadir$^*$\\
  UFR Mathematics and Applications\\
  Faculty of Science and Technology\\
  University of Hassan II-Mohammedia \\
  B.P. 146, Mohammedia,
  Morocco (2008)\\
  Email: khadir@hotmail.com
}

\date{}
\begin{document}
\maketitle

 \baselineskip=20pt

\noindent\hrulefill

{\abstract \small In this paper we present a new efficient algorithm for factoring the RSA and the
Rabin moduli in the particular case when the difference between their two prime factors
is bounded. As an extension, we also give some theoretical results on factoring integers.}

\noindent\hrulefill

 \vspace{0.3cm} \noindent {\small \bf Keywords :} \
{\small RSA, Rabin cryptosystem, Factorization problem.}

\vspace{0.5cm}

\noindent {{\small \bf MSC 2010 : } {\small 11B50 94A60}

 \section{Introduction}
The security of the RSA [11], and Rabin [10] cryptosystems is based
on the hardness of factoring integers. The secret keys can be founded if we
succeed in decomposing the modulus which is the product of two large
prime factors.

Many authors have addressed the problem and currently the fastest
known algorithms are Elliptic Curves Method [5], and Number Field
Sieve [2]. In an exercise, Stinson [12], has evoked the
possibility of factoring the RSA modulus if the two factors are
too close. In 1999, Boneh and al. [1] described a polynomial time
algorithm for factoring $n =p^rq$ when the exponents $r$ is large.
More recently, in 2007, Coron an May [3] presented the first
deterministic algorithm for factoring the RSA modulus in
polynomial time but they used the public and the secret key pair
$(e,d)$. Our work consists on giving a simple algorithm for
factoring the RSA and the Rabin moduli in the particular case when
the difference between the two prime factors is less than
$2^{\frac{k+5}{4}}$ where $k$ is the bit-size of the modulus. The
paper is organised as follow: Section 2 is devoted to our main
result. In section 3 we discuss an extension but only in its
theoretical aspect. We conclude in Section 4. Throughout the
paper, we shall use standard notation. In particular N is the set
of all natural integers $0,1,2,3, \dots$ and
$\mathbb{N}^*=\mathbb{N}-\{0\}$. The largest integer which does
not exceed the real x is denoted by $\lfloor x \rfloor$. It is
also the integer part and the floor of $x$. Thus we have $\lfloor
x\rfloor \leq x < \lfloor x\rfloor+ 1$. The bit-size of a positive
integer $n$ is the number of bits in its binary representation.
So, the bit-size of $n$ is $k$, $\Leftrightarrow$ $\displaystyle
n=\sum_{i=0}^{k-1} 2^ia_i$ with every $a_i \in \{0,1\}$ and
$a_{k-1}=1$.

\section{Main Results}
We begin with a lemma that we shall use in the proof of our main
theorem.

\vspace{0.3cm}  \noindent {\bf Lemma 2.1} Let $n,m$ be two elements of $\mathbb{N}^*$ and let $\alpha_{n,m}$ denotes the number of
perfect squares $x^2$ such that $n<x^2 \leq m$. Then we have: $\displaystyle \alpha_{n,m}<\frac{m-n}{\sqrt{n}+\sqrt{m}}+1$.

\proof Consider the set $E_n=\{x \in \mathbb{N} \ | \ x^2\leq n \}$. Since $E_n$ is also $\{x \in \mathbb{N}\ | \ x \leq \sqrt{n}\}$, its cardinality is $\lfloor \sqrt{n} \rfloor+1$ and then $\alpha_{n,m}=\lfloor \sqrt{m} \rfloor-\lfloor \sqrt{n} \rfloor$. If we put $k=\lfloor \sqrt{n} \rfloor$ and $l=\lfloor \sqrt{m} \rfloor$, which means that $k\leq \sqrt{n}<k+1$ and $l\leq \sqrt{m}<l+1$, we obtain $l\leq \sqrt{m}$ and $-k <1-\sqrt{n}$. Hence $\displaystyle \alpha_{n,m}=l-k<\sqrt{m}-\sqrt{n}+1=\frac{m-n}{\sqrt{n}+\sqrt{m}}+1$.

\vspace{0.5cm}

Now we can move to the main theorem which allows us to compute
efficiently the two prime factors $p$ and $q$ of an RSA or a Rabin
modulus in a particular case. The proof of this theorem relies on
the last lemma.

\vspace{0.3cm}  \noindent {\bf Theorem 2.2} Let $n$ be the modulus
of an RSA or a Rabin cryptosystem whose bit-size is denoted by $k
\in \mathbb{N}^*$. If its two prime factors $p$ and $q$ satisfy
the inequality $|p-q| \leq 2^{\frac{k+5}{4}}$, then we can compute
them efficiently

\proof First notice that the hypothesis of our theorem can exist
in practice: for example when $p$ and $q$ are twin primes. Without
loss of generality we can assume that $2 < p < q$. As the factors
$p$ and $q$ are odd, we put $q=p+2i$ where $i \in \mathbb{N}$.
Since $n=pq \Leftrightarrow n+i^2=(p+i)^2$, the integer $n+i^2$ is
a perfect square bounded below by $n$ and above by
$n+2^{\frac{k+1}{2}}$ because $2i = q-p\leq2^{\frac{k+1}{2}}
\Rightarrow i^2 \leq 2^{\frac{k+1}{2}}$. Let $m = n + i^2$. By the
last lemma, the number $\alpha_{n,m}$ of perfect squares between
$n$ and $m$ is satisfying the inequality $\displaystyle
\alpha_{n,m}<\frac{i^2}{\sqrt{n+i^2}+\sqrt{n}}+1$. We then deduce
that $\displaystyle \alpha_{n,m}<\frac{i^2}{2\sqrt{n}}+1$ and as
$k$ is the bits-size of $n$, that $\displaystyle \alpha_{n,m} <
\frac{i^2}{2.2^{\frac{k-1}{2}}}+1$. Hence $\displaystyle
\alpha_{n,m} < \frac{2^{\frac{k+1}{2}}}{2^{\frac{k+1}{2}}}+1$

But $\alpha_{n,m}$ is a natural integer so $\alpha_{n,m}=1$. This
means that $n +i^2 = (p +i)^2$ is the only perfect square between
$n$ and $n + 2^{\frac{k+1}{2}}$. That is also the first perfect
square $n_0^2$ greater than $n$ and so $n_0 = \lfloor \sqrt{n}
\rfloor +1$. This allows us to compute the factors $p$ and $q$ :
$n + i^2 = n_0^2 \Longrightarrow n = (n_0+i)(n_0 - i)$
$\Longrightarrow$ $p =n_0-i$ and $q=n_0+i$. This theorem leads to
the following algorithm where comments are delimited by braces.

\vspace{0.5cm}

\noindent
\textbf{Algorithm} \\
\textbf{Input:} A modulus $n > 0$ with $|p-q| \leq 2^{\frac{k+5}{4}}$. \\
\textbf{Output:} The two prime factors $p$ and $q$. \\
\hspace*{0.2cm} (1) $n_0  \leftarrow \lfloor \sqrt{n} \rfloor +1$ \{ $n_0$ is the first integer square $>$ n \} \\
\hspace*{0.2cm} (2) $I \leftarrow n_0^2-n$ \{ $I$ is an intermediate variable \} \\
\hspace*{0.2cm} (3) $i \leftarrow \sqrt{I}$ \{ $i^2$ is a perfect square \} \\
\hspace*{0.2cm} (4) $p  \leftarrow n_0 - i$ \{ We compute $p$ and $q$ \} \\
\hspace*{0.2cm}  (5) $q   \leftarrow n_0 + i$ \\
\hspace*{0.2cm} (6) Output $p$ and $q$. \\

\vspace{0.3cm}  \noindent {\bf Example 2.3} Let try the method on
the mythic example given by the authors of the RSA cryptosystem
[11]. They took $n = 2773$, $p = 47$ and $q = 59$. With the
algorithm above we retrieve easily the two prime factors. Indeed
the first integer square greater than $n$ is $n_0^2=(\lfloor
\sqrt{n} \rfloor +1)^2 =53^2 = 2809$, so $n_0^2-n=36 = 6^2 = i^2$
and then $p = n_0-i = 47$ and $q = n_0 + i = 53$. Let check that
$p$ and $q$ satisfy the condition in the theorem, $n = 2773$ has
$k = 12$ bits in its binary representation, thus
$2^{\frac{k+5}{2}}=2^4\sqrt[4]{2} \Rightarrow |q-p|=12 \leq
2^{\frac{k+5}{2}}$.

\vspace{0.5cm}

On an other hand there exist integers for which we cannot apply
the theorem. Take for example $n = 1081$. The first, integer
square greater than $n$ is $n_0^2= 1089$, but $n_0^2-n = 8$ is not
a perfect square. Here the hypothesis is not valid with the values
$p = 23$, $q = 47$ and $k = 11$. Observe that when our method
fails, it gives information on the two factors $p$ and $q$, namely
that they are not very close to each other. From the theorem we
deduce that some integers should be avoided as RSA or Rabin
moduli. More precisely:


 \vspace{0.3cm}  \noindent {\bf Corollary 2.4.} Let $n = pq$, $p, q > 2$ be the modulus of an RSA or a Rabin
cryptosystem which bit-size is denoted by $k \in \mathbb{N}^{*}$. Assume that $p$ and $q$ have the
same bit-size $\displaystyle \frac{k}{2}$. If $p$ matches $q$ on the $\displaystyle \frac{k}{4}$
most significant bits, then we can compute
the two prime factors $p$ and $q$.

\proof We have in this situation: $|q-p| \leq 2^{\frac{k}{4}} \leq
2^{\frac{k+5}{4}}$

\section{Extension of the method}

The purpose of this section is to generalize our method. The extension
is mainly of theoretical interest. However we can compute factors by
"factoring with a hint" [1], [2] or the help of an oracle. The following
proposition shows that, when $n = pq$ is the product of two unknown
prime factors, if we can find a prime number $r$ such that $rp$ is close to
$q$, and therefore $rn$ is close to a perfect square, then we can compute $p$ and
$q$. The difficulty of factoring $n$ directly, is transformed into the difficulty
of computing this coefficient $r$. When this situation occurs, since $r$ is an
integer, the factors $p$ and $q$ must be unbalanced [6]. It seems that, in this
case, classical algorithms are not very efficient.

\vspace{0.3cm}  \noindent {\bf Proposition 3.1} Let $n \in
\mathbb{N}^{*}$ be the product of two prime factors $p$ and $q$,
$2 < p < q$. If we can compute efficiently an odd integer $r > 2$
such that $|q - rp| \leq  2^{\frac{K+5}{4}}$, where $K$ is the
bit-size of the integer $rn$, then we can compute the factors $p$
and $q$.

\proof We put $N = rn$, $P =rp$ and $Q = q$. So $N = PQ$ and as
$P$ and $Q$ are odd we assume that $Q > P$, and $Q = P + 2I$.
Using a technique like that in the proof of Theorem 2.2 but with
the new parameters $N, P, Q, K, I$ instead of $n, p, q, k, i$, we
show that there is only one perfect square between N and $N +
2^{\frac{K+1}{2}}$ and it is the first square $N_0^2$ greater than
N. We have also: $N = N_0^2-I^2=(N_0-I)(N_0 + I)$. We wish to have
$p$ as a factor of $N_0 -I$ and $q$ as a factor of $N_0 + I$.
Indeed, suppose that $r = r_1r_2$with $N_0 -I = r_1$ and $N_0 + I
= r_2 pq$. We have: $N_0 - I = r_1$ and $N_0 + I = r_2 pq
\Rightarrow 2I = r_2 pq-r_1 \Rightarrow q -rp = r_2 pq -r_1
\Rightarrow r_1 -rp = r_2 pq - q$.

This leads to a contradiction since $r_1 - rp < 0$ and $r_2 pq - q > 0$. We
conclude that $p$ is a factor of $N_0 - I$ and $q$ is a factor of $N_0 + I$ and then we
can compute them

\vspace{0.3cm}  \noindent {\bf Example 3.2} Let $n = 15211$ $(= 41 \times 371)$.
If we take $r = 9$, the first square $N_0^2$ greater than $rn = 136899$ is $370^2$. So
$N_0^2- rn = 1$ and therefore $rn = 369 \times 371$. By looking first for the factors
of the artificial coefficient $r = 9$ we easily retrieve that $p = 71$ and $q = 371$.

\vspace{0.5cm}

If the factors $p$ and $q$ are balanced which is the case in
standard RSA and Rabin cryptosystems [6], we have the result:

\vspace{0.3cm}  \noindent {\bf Proposition 3.3} Let $n \in \mathbb{N}^{*}$ be the product of two prime factors $p$ and $q$,
$2 < p < q$. If we can compute efficiently two odd integers $r, s$ such that $s < p$
and $|sq - rp| \leq 2^{\frac{k+5}{4}}$ where $K$ is the bit-size of the integer $rsn$, then we can
compute the factors $p$ and $q$.

\proof For simplicity we suppose that $sq > rp$. The same
argumentation as in the proof of Proposition 3.1 shows that the
first perfect square $N_0^2$ greater than $rsn$, verify $N_0^2
-rsn = I^2$ where $2I = sq - rp$. So $rsn =(N_0 - I)(N_0 + I)$.
From this decomposition let show that $p$ is a factor of $N_0 - I$
and $q$ a factor of $N_0 + I$ and then it's easy to compute them.
Suppose that we have $rs = uv$ with $N_0 - I = u$ and $N_0 + I =
vpq$. We then have $2I = vpq - u$ and thus $sq - rp = vpq - u$.
This leads to $ u - rp = vpq - sq$. But $vpq-sq = q(vp- s)$ is
positive; and $v(u-rp) = uv-vrp = r(s-vp)$ is negative. Hence $pq$
cannot divide $N_0 - I$ and therefore $p$ is a factor of $N_0 - I$
and $q$ factor of $N_*0 + I$.

\vspace{0.3cm}  \noindent {\bf Example 3.4} Let $n = 24961$  $(= 109 \times 229)$ as in an example from [7].
Here we cannot apply Theorem 2.2. If we take $r = 23$ and $s = 11$, the first
square $N_0^2$ greater than $rsn = 3569423$ is $2513^2$, and $N_0^2-rsn = 62$. So
$rsn = 2507 \times 2519$ and by decomposing each factor we retrieve $p = 109$
and $q = 229$.

\vspace{0.5cm}

Our theoretical method can be extended in order to be applied for
factoring any integer $n$. By the fundamental theorem of arithmetic, every
positive integer $n$ can be written as a product of primes. So it n can be
made in the form $n = f g$, where $f$ and $g$ are two factors not necessary
prime. If for one couple $( f , g)$ the difference $|g-f|$ is not very large, then
we can compute $f$ and $g$.

\vspace{0.3cm}  \noindent {\bf Proposition 3.5.} Let $n \in \mathbb{N}^{*}$ be the product of two odd factors $f$ and $g$,
$2 < f < g$. If we have $|g-f| \leq 2^{\frac{k+5}{4}}$
where $k$ is the bit-size of the integer $n$, then we can compute the factors $f$ and $g$.

\proof Similar to the proof of Theorem 2.2.

\vspace{0.3cm}  \noindent {\bf Example 3.6.} Let $n = 155227$ $(= 17 \times 23 \times 397)$.

The first perfect square $n_0^2$ greater than $n$ is $394^2$ and $n_0^2-n=3^2$. So $n = 391 \times 397$.

There is an other interesting example with $m = 2^{4\alpha+2} + 1,
\alpha \in \mathbb{N}$. (see [4] for $\alpha = 53$). The exponent
is simply the double of odd integers. The first perfect square
grater than $m$ is $m_0^2= (2^{2\alpha+1} + 1)^2$. So $m_0^2-m =
(2^{\alpha+1})^2$, and therefore $p = m_0 - 2^{\alpha+1} =
2^{2\alpha+1} - 2^{\alpha+1} + 1$ and $q = m_0 +2^{\alpha+1}
=2^{2\alpha+1} + 2^{\alpha+1} + 1$. Observe that $m$ is also a
multiple of $5$.

\vspace{0.5cm}

The last result in this paper concerns integers $n$ for which no couple
of factors $( f , g)$ verify $|g-f| \leq 2^{\frac{k+5}{4}}$. In this case we use a coefficient $r$ to
correct the situation and work on the new integer $rn$ before coming back
to $n$ and compute efficiently one of its factors. We formulate the idea in the
next theorem:

\vspace{0.3cm}  \noindent {\bf Theorem 3.7.} Let $n \in \mathbb{N}^{*}$ be an odd integer. Assume that we can compute efficiently
an odd integer $r$ such that $rn$ becomes the product of two factors $f$ and $g$
such that $r < f$ (or $r < g$) and $|g-f| \leq 2^{\frac{k+5}{4}}$ , where $K$ is the bit-size of the
integer $rn$, then we can compute a factor of $n$.
\proof Similar to the proof of Corollary 2.4.

\vspace{0.3cm}  \noindent {\bf Example 3.8.} Let $n = 136793$ $(= 29 \times 53 \times 89)$.

\noindent Here we cannot apply Proposition 3.5. With $r = 17$ or ($r = 49$) we have
$rn = 2325481$. The first perfect square $N_0^2$ greater than $rn$ is $1525^2$ and
$N_0^2 - rn - 12^2$. So $rn = 1513 \times 1537$ and by looking for the artificial
coefficient $r$ we find two factors of $n$ namely $f = 89$ and $g = 1537$

\section{Conclusion}

We have described a, algorithm for factoring the RSA and the Rabin
moduli in a particular case. This class of integers should be avoided in
cryptographic applications. The algorithm does not use divisions.We need
in the future to ameliorate the bound $2^{\frac{k+5}{4}}$, in order to include more prime
factors.

Furthermore, we have also discussed new ideas about integer factorization.
The technique is only theoretical but we believe that it can lead to
efficient algorithms for some classes of integers. We underline that in the
case of the RSA cryptosystem, we did not use the knowledge of the public
and secret key pair $(e, d)$.


\end{document}